\documentstyle[12pt]{article}
\begin{document}
\baselineskip24pt
\begin{center}
\begin{Large}
{\bf  Non-resonant microwave absorption studies of superconducting MgB$_2$}
\end{Large}

\vskip 1cm
{\large Janhavi P. Joshi, Subhasis Sarangi, A. K. Sood and S. V. Bhat\\
 Department of Physics,  Indian Institute of
 Science, Bangalore-560012, India}\\
{\large and \\ Dilip Pal\\
Tata Institute of Fundamental Research, Mumbai, 400005, India
}
\vskip1cm
\end{center}

Non-resonant microwave absorption(NRMA) studies of superconducting MgB$_2$ at a frequency of 9.43 GHz in the field range -50 Gauss to 5000 Gauss are reported. The NRMA results indicate near absence of intergranular weak links. A linear temperature dependence of the lower critical field H$_c1$ is observed indicating a non s-wave superconductivity. However, the  phase reversal of the NRMA signal which could suggest d-wave symmetry is also not observed. 

\newpage

The newly discovered \cite{akim} intermetallic compound superconductor MgB$_2$ has given rise to a flurry of scientific activity as well as its own share of controversies. While the observation of boron isotope effect in MgB$_2$ indicated it to be a phonon mediated BCS superconductor\cite{bud}, the temperature dependence of the lower critical field H$_{c1}$ was found to be linear\cite{wen} expected for a d-wave symmetry of the order parameter. Another pleasant surprise has been the reported absence of weaklink structure between intergranular contacts in the polycrystalline samples, leading to high critical current density (J$_c$) even in the presence of a magnetic field.\cite{larb,kim} This  result, which is in quite a contrast to the high T$_c$ superconductors(HTSC), that abound in intergranular weak links,  makes MgB$_2$ a very good potential candidate for practical applications. 

The technique of non-resonant microwave absorption (NRMA) has proven to be a valuable tool in detecting and characterising superconductivity in the case of HTSC\cite{svb1}. In this technique, the sample is studied using a continuous wave Electron Paramagneic Resonance (EPR)  spectrometer, by recording the magnetic field dependence of the power absorption. In standard EPR spectrometers, the magnetic field is usually modulated at 100 kHz and phase sensitive detection is used to record the field derivative dP/dH of the power absorbed. In the case of HTSC two processes mainly contribute to this field dependent dissipation. One, the decoupling of the Josephson junctions (JJs) due to the magnetic field\cite{dulcic} and the other the Lorentz force driven motion of the quantised flux lines\cite{portis}, the so called fluxons. In all but the extremely well prepared thin film and single crystalline HTSC samples the former contribution dominates, especially at low fields making dP/dH {\it vs} H look like a regular EPR signal except for the reversal of the phase. The latter is a consequence of the loss in the mixed state of the superconductors being an increasing function of the field with a minimum at zero field.  It turns out that the two contributions can be easily distinguished from each other from the characteristic field dependences, reflected in the line shapes\cite{vvs}. Thus, the decoupling of the JJs leads to a narrow derivative looking signal centred around zero field while the fluxon motion results in a  quasi-linear dependence on field and therefore the dP/dH appears to have near field independence. In this report, we examine superconducting MgB$_2$ samples with a view to charactarise them using the technique of NRMA. 

The sample of MgB$_2$ used was obtained from M/s Alpha Aesar of John Matthey GmBH, Germany and was used without further processing except for pelletising under a pressure of 2 tons. AC susceptibility measurements indicated a T$_c$ of 39K. The NRMA signals were recorded using a Bruker 200D X band (microwave frequency: 9.43 GHz) EPR spectrometer equipped with an Oxford Instruments ESR 900 continuous flow cryostat, in the temperature range 4~K to room temperature. A typical modulation amplitude of 4 Gauss peak to peak  and a microwave power of 20 mW were used. The sample was zero field cooled to 4~K and the signals were recorded between -50 and 5000 Gauss for both forward and reverse  scans of the field during both cooling and warming of the sample. 

Fig. 1 shows  typical NRMA signals recorded at a few representative temperatures recorded in the warming run. The signals recorded while cooling (not shown)are essentially similar. In the normal state only a very weak EPR signal due to some impurity in the cavity is observed. Below T$_c$ ($\sim$ 39~K) the NRMA signal appears centred at zero field which on cooling increases both in intensity and extent in the field. By 34~K it is seen to have extended beyond the limits of our magnetic field (0.5 Tesla). At zero field, a sudden rise in the signal (note its opposite phase to that of the EPR signal as expected for a normal NRMA signal) is observed and at a temperature dependent higher field there is a change in the slope of the signal. The point at which the slope change is observed (as marked by the arrows on a few signals) can be associated with the lower critical field. We denote this field by H$^*_{c1}$ to indicate that it is  uncorrected for demagnetisation factor and geometric and/or surface barriers.  Figure 2 shows the temperature dependence of H$^*_{c1}$. It is seen that the temperature dependence is approximately linear, attributed to a non s-wave superconductor\cite{wen}.  Such linear dependences have also been observed by other workers\cite{joshi}. The opposite scans of the magnetic field bring out the rectangular nature of the hysteresis loops similar to the ones observed in single crystal HTSC at low temperatures  indicative of the large critical current densities.

The crucial difference between these NRMA signals and those observed in polycrystalline samples of HTSC is the near absence of narrow derivative looking signal observed in the latter at low fields commonly attributed to JJs. This might indicate the absence of intergranular weak links. However, the presence of a significant field dependent dissipation below H$_{c1}$ (indicated by  ellipses in fig. 1) obtained in MgB$_2$ needs to be explained. While quasi-particle excitation is a possible candidate, we believe the temperature and field dependences of the observed signals, do not favor this mechanism. NRMA technique being very sensitive to the JJs,  can pick up even minute contributions from them. Therefore, we attribute the signal which is observed below H$_{c1}$ to the presence of a few weak  JJs in the sample. The temperature dependence of the signal below H$_{c1}$ also supports this possibility. 

Another interesting feature of the NRMA signals is the absence of phase reversal of the signal as a function of temperature. Such a phase reversal has been observed in almost all HTSC  materials \cite{bhat} especially in their granular form; one of the causes is understood to be the paramagnetic Meissner effect(PME) arising due to d-wave symmetry of the order parameter\cite{wohl} and/or the presence of $\pi$ junctions \cite{rice,bhat2}. However, inspite of varying microwave power and  modulation amplitude, we were unable to observe any such phase reversal in the NRMA signals from MgB$_2$. It appears therefore, that the linear temperature dependence of H$_c1$ and the absencce of the phase reversal in NRMA signals need to be examined and understood further with regard to their implications for the symmetry of the order parameter. 

In conclusion, we have studied the novel superconductor MgB$_2$ using the technique of NRMA. The NRMA signals from MgB$_2$ are very different from that of HTSC. Low field derivative like narrow signals characteristic of the Josephson junctions is absent while the rectangular hysteresis loops indicative of large critical current densities are observed. The lower critical field is observed to decrease linearly with temperature.  The absence of phase reversal of the NRMA signal is noted as another unique feature of this superconductor, especially because such a phase reversal can point towards the d-wave symmetry of the order parameter through PME.

\begin{large}
\noindent Acknowledgments
\end{large}\\
The authors acknowledge  Professors Arun Grover and S. Ramakrishnan of TIFR, Bombay for providing the samples. 
JJ  would like to thank  CSIR, India, AKS and SVB acknowledge DST and UGC, India for financial support.

\newpage 
\begin{large}
Figure Captions
\end{large}

\noindent {\bf Figure 1}\\
The typical NRMA signals recorded during the warming of the sample at a few representative temperatures for both forward and reverse scans of the magnetic field. The dotted ellipse indicates the part of the signal tentatively attributed  to a very small number of weak links forming JJs in the sample. The vertical arrows mark the field at which the slope changes indicating the onset of the entry of the magnetic field in the form of vortices in the sample.\\

\noindent{\bf Figure 2}\\
The H$^*_{c1}$ obtained from the slope change in the NRMA signals and uncorrected for demagnetisation factors geometrical and/or surface barrier effects,  plotted as a function of reduced temperature (T/T$_c$), showing quasi-linear behaviour.  


\newpage
\begin{thebibliography}{25}
\bibitem{akim} J. Akimitsu, Symposium on Transition Metal Oxides, Sendai, Japan, January 10, 2001.
J. Nagamatsu, N. Nakagawa, T. Murunaka, Y, Zenitani and Y. Akimatsu, Nature, {\bf 410}, 63, (2001).
\bibitem{bud}S. L. Bud'ko, G. Lapertot, C. Petrovic, C. E. Cunningham, N. Anderson and P. C. Canfield, Phys. Rev. Lett. {\bf 86}, 1877, (2001).
\bibitem{wen} S. L. Li, H. H. Wen, Z. W. Zhao, Y. M. Ni, Z. A. Ren, G. C. Che, H.P.Yang, Z.Y.Liu and Z. X. Zhao, cond-mat/0103032  (2001).
\bibitem{larb}D. C. Larbalestier, L. D. Colley, M. O. Rikel, A. A. Polyanskii, J. Jiang, S. Patnaik, X. Y. Cai, D. M. Feldmann, A. Gurevich, A. A. Squitieri, M. T. Naus, C. B. Eom, E. E. Hellstorm, R. J. Cava, K. A. Regan, N. Rogado, M. A. Hayward, T. He, J. S. Slusky, P. Khalifah, K. Inumaru and M. Haas, Nature {\bf 410}, 186 (2001).
\bibitem{kim}Kijoon H. P. Kim, W. N. Kang, Mun-Seog Kim, C. U. Jung, Hyeong-Jin Kim, Eun-Mi Choi, Min-Seok Park and Sung-Ik Lee, COnd-mat/0203176  
\bibitem{svb1} S. V. Bhat, P. Ganguly and C. N. R. Rao, Pramana J. Phys. {\bf 28} (1987).
Blazey K. W., Muller K. A., Bednorz J. G., Berlinger W., Amoretti G. Buluggiu E., Vera A. and Matacotta F. C., Phys. Rev. {\bf B 36}, 7241 (1987).
\bibitem{dulcic} A. Dulcic, B. Rakvin, M. Pozek, Europhys. Lett.{\bf 10} 593, 
(1989).
\bibitem{portis} A. M. Portis, K. W. Blazey, K. A. Muller and J. G. Bednorz, Europhys. Lett. {\bf 5}, 467-472 (1988).
\bibitem{vvs} V. V. Srinivasu, B. Thomas, M. S. Hegde and S. V. Bhat, J. Appl. Phys. {\bf 75} 4131 (1994).
\bibitem{joshi}A. G. Joshi, C. G. S. Pillai, P. Raj and S. K. Mallik, cond-mat/0103302 (2001).
\bibitem{bhat}S. V. Bhat in {\it Studies of High Temperature Superconductors}, ed. by A. V.
Narlikar,(Nova Science Publishers, New York, 1996), Vol.18, pp. 241-259.
\bibitem{wohl}  W. Braunisch, N. Knauf, V. Kataev, S. Neuhausen, A. Gruitz, A. Kock, B. Roden, D. Khomskii and D. Wohlleben, Phys. Rev. Lett. {\bf 68}, 1908 (1992).
\bibitem{rice} M. Sigrist and T. M. Rice, J. Phys. Soc. Jpn {\bf 61}, 4283 (1992).
\bibitem{bhat2} S. V. Bhat, A. Rastogi, N. Kumar, R. Nagarajan and C. N. R. Rao, Physica {\bf c 219}, 87 (1994).

\end{thebibliography}
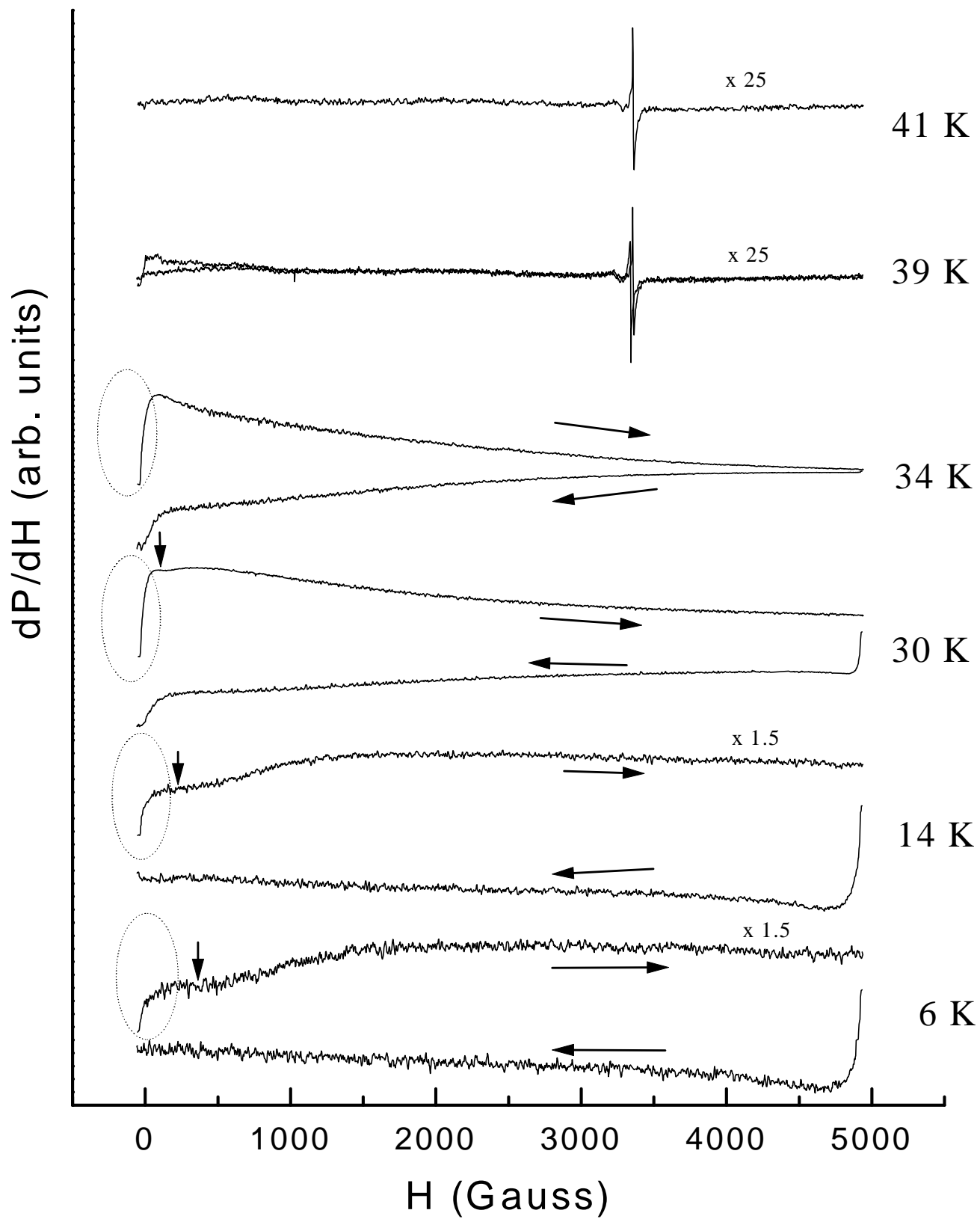
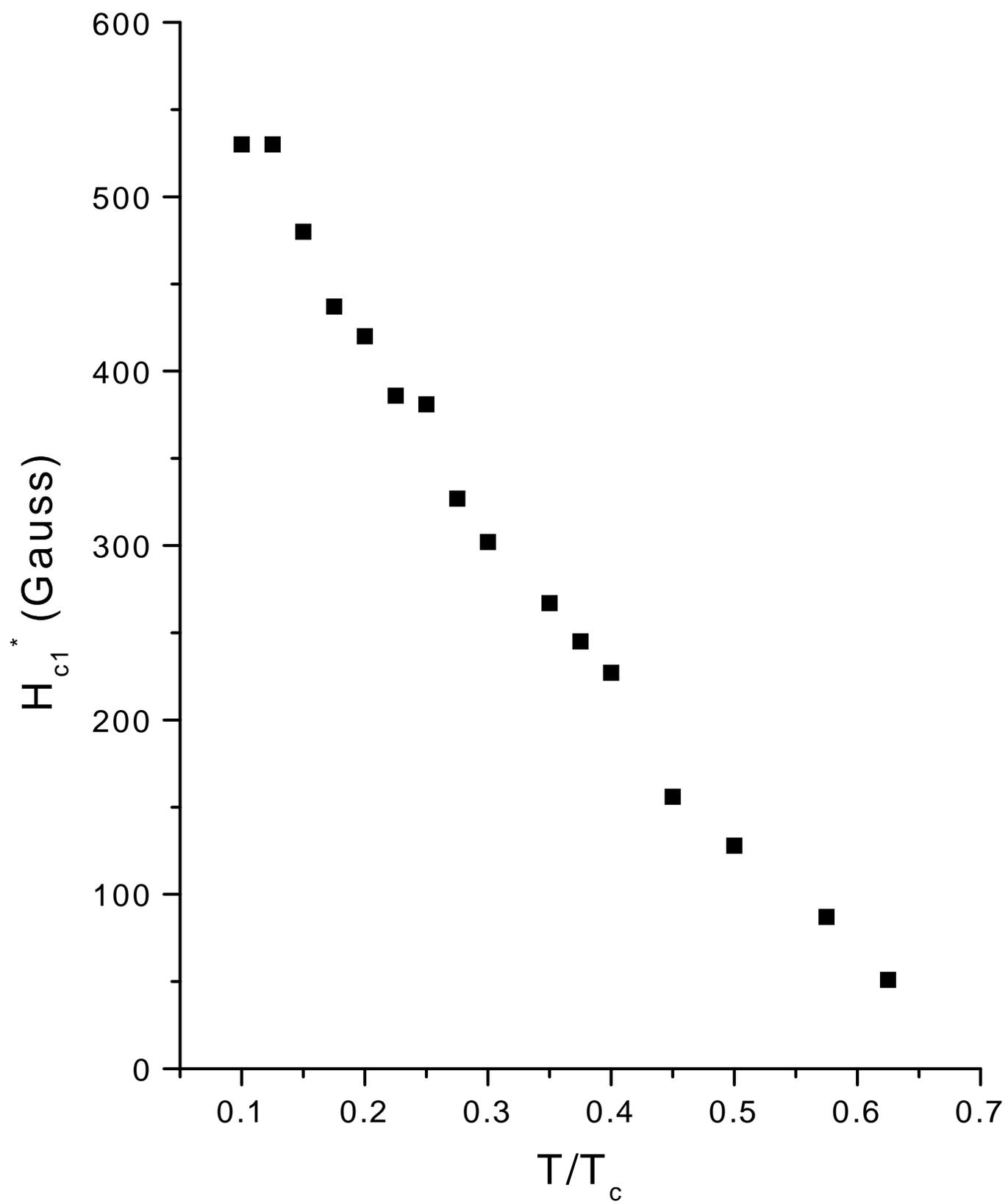
\end{document}